# How do digits emerge? - Mathematical Models of Limb Development

Dagmar Iber[1,2,*], Philipp Germann[1]

**Affiliations:**
1 - Department of Biosystems, Science and Engineering (D-BSSE), ETH Zurich, Mattenstraße 26, 4058 Basel, Switzerland
2 - Swiss Institute of Bioinformatics (SIB), Switzerland

* Corresponding Author and Contact Person for Reprint Requests:

Dagmar Iber

Department for Biosystems, Science, and Engineering (D-BSSE),

ETH Zurich

Mattenstraße 26

4058 Basel

Switzerland

Tel + 41 61 387 3210

Fax + 41 61 387 31 94

dagmar.iber@bsse.ethz.ch

Keywords: Limb Development, Endochondral Ossification, Turing Pattern, Computational Biology

**ABSTRACT (146)**

The mechanism that controls digit formation has long intrigued developmental and theoretical biologists, and many different models and mechanisms have been proposed. Here we review models of limb development with a specific focus on digit and long bone formation. Decades of experiments have revealed the basic signalling circuits that control limb development, and recent advances in imaging and molecular technologies provide us with unprecedented spatial detail and a broader view on the regulatory networks. Computational approaches are important to integrate the available information into a consistent framework that will allow us to achieve a deeper level of understanding and that will help with the future planning and interpretation of complex experiments, paving the way to *in silico* genetics. Previous models of development had to be focused on very few, simple regulatory interactions. Algorithmic developments and increasing computing power now enable the generation and validation of increasingly realistic models that can be used to test old theories and uncover new mechanisms.

# Introduction

Limb development has long served as a model system for organogenesis before methods became available that facilitated the analysis of the development of vital organs (Zeller et al., 2009). As a result, an unmatched level of molecular detail has been defined, which makes the limb a well-suited system to develop detailed mathematical modelling approaches (Iber and Zeller, 2012). The first computational models of limb development were concerned with the growth of the limb bud and suggested that growth had to be anisotropic to yield the observed embryonic limb shapes (Ede and Law, 1969). Recent 3D imaging of the limb in combination with an earlier Navier-Stokes based growth model (Dillon and Othmer, 1999) indeed supports the notion of anisotropic growth in the limb (Boehm et al., 2010).

Most modelling attention has focused on the emergence of digits during limb development (Maini and Solursh, 1991). Nascent digit condensations become first visible around embryonic day (E) 11.5 and express *Sox9* (Figure 1A) (Kawakami et al., 2005). Interestingly, the digits appear neither simultaneously nor in a strict posterior to anterior sequence (or vice-versa) (Figure 1B) (Zhu et al., 2008). The mechanism that controls digit formation has long intrigued developmental and theoretical biologists, and in this review we will focus on proposed mechanisms and models, starting with the earliest approaches (French Flag model and Turing mechanism) and extending it to more recent approaches that take the detailed regulatory interactions and the physiological, growing limb bud geometry in the developing limb bud into account. Finally, we discuss models for the formation of long bones, once the digit condensations have been defined.

# The French Flag Model

Grafting experiments in chicken embryos showed that a small piece of tissue from the posterior part of the limb bud, the so-called zone of polarizing activity (ZPA), would induce a second, mirror-symmetric set of digits when grafted to the anterior side of a recipient limb bud (Saunders and Gasseling, 1968), as reviewed in (Tabin, 1991). Wolpert proposed that the ZPA was the source of a diffusible morphogen, and that a diffusion-based gradient along the anterior-posterior (AP) limb axis would induce the different digit identities (Wolpert, 1969). According to this so-called French Flag model cells would respond differently above and below certain concentration thresholds along the gradient (Figure 2A) (Wolpert, 1969). The French flag model was supported by further experiments that showed that smaller grafts (producing presumably less morphogen) induced only partial duplications (Tickle, 1981) (Figure 2B). About 25 years later the proposed morphogen-type signal was identified as Sonic Hedgehog (SHH) (Riddle et al., 1993). In fact, genetic and molecular analysis established that SHH fulfilled several of the criteria of a true morphogen. Thus inactivation caused the loss of all, but the most anterior digit (thumb) (Chiang et al., 1996; Kraus et al., 2001), and the forced expression of *Shh* in fibroblasts converted these into cells with polarizing activity such that their implantation into the anterior limb bud induced formation of a second, mirror-symmetric set of digits in a concentration-dependent fashion (Riddle et al., 1993; Yang et al., 1997).

Later experiments showed that it was important to distinguish between mechanisms that would induce the emergence of digits and those that would specify the digit type. Mouse limb buds lacking both *Shh* and the downstream mediator *Gli3* have a polydactylous phenotype with up to ten digits (Litingtung et al., 2002; te Welscher et al., 2002), such that SHH signalling is clearly not necessary for digits to emerge. The digits, however, all looked the same, such that SHH appears to be necessary to specify the digit type. Furthermore, digits appear neither simultaneously nor in strict posterior to anterior sequence (or vice-versa)

(Figure 1B) (Zhu et al., 2008). Thus, the formation of digits cannot be explained by a simple spatial SHH morphogen concentration gradient acting across the limb field. As regards to digit specification, there is evidence that long-range SHH signalling contributes to patterning of the middle and anterior digits 3 and 2, but specification of the most posterior digits 4 and 5 has been suggested to depend on the length of exposure of progenitors to SHH signalling (Harfe et al., 2004).

However, there are also limitations to a simple SHH-based model for digit specification, including the robustness of the process. The expression of *Shh* has been shown to fluctuate during limb bud development (Amano et al., 2009), and neither removal of one copy of *Shh* (Bénazet et al., 2009; Chiang et al., 1996), nor posterior implants of *Shh* expressing cells alter the digit pattern (Riddle et al., 1993). Theoretical considerations suggest that for a single morphogen-threshold-based mechanism even small changes in concentrations at the source would shift the position at which the pattern would emerge (Figure 2C) (Lander et al., 2009). In addition to noise at the source, there will also be variation in ligand transport, degradation, and receptor binding (Bollenbach et al., 2008). While spatial and temporal averaging of ligand concentrations can enhance the precision of a morphogen read-out (Gregor et al., 2007), and feedbacks may exist to buffer changes in the SHH concentration, a more sophisticated regulatory network must be invoked to explain robust pattern formation in the limb. In fact, experiments show that while the expression of *Shh* is reduced to 65-70% of its normal value, the expression of SHH- and BMP-dependent genes (*Gli1* and *Msx2* respectively) is normal in *Shh* heterozygous mice (Bénazet et al., 2009).

There is the added problem of size. Chicken limb buds are significantly larger than mouse limb buds at early stages, but the patterning mechanisms appear overall highly conserved between species, as also demonstrated by cross-species grafts (Tabin, 1991; Tickle et al., 1976). As illustrated in Figure 2D, any pattern arising due to threshold values set by a simple linear or exponential diffusion gradient would not scale with the size of the domain. In particular, the blue and white stripes of the French flag in the larger domain would be equal to the ones of the smaller domain, while the red stripe would expand (i.e no scaling). The problem was previously recognized for the Bicoid gradient in differently sized Drosophila embryos (Gregor et al., 2007), as well as in many other developmental systems (Ben-Zvi et al., 2008; Umulis et al., 2010). While several mechanisms have been proposed to deal with the problem of correct scaling (Ben-Zvi et al., 2011; Ben-Zvi et al., 2008; Lauschke et al., 2013; Umulis, 2009), these have remained controversial or would not apply to the limb.

## Turing Pattern

In his seminal essay in 1952, long before the sequence and structure of the underlying gene products became known, Turing proposed that "chemical substances, called morphogens, reacting together and diffusing through a tissue, are adequate to account for the main phenomena of morphogenesis" (Turing, 1952). Alan Turing showed that two components, which diffuse at different speeds and regulate each other in a specific manner, can give rise to a wide range of different patterns. The details of Turing's theory have fascinated generations of biologists and they have repeatedly been reviewed (Kondo and Miura, 2010); Murray's classical textbook on Mathematical Biology gives details on the mathematical aspects (Murray, 2003)). The Turing mechanism is based on a diffusion-driven instability and is sufficiently flexible to reproduce virtually any pattern as long as the reactions and parameters are appropriately adjusted (Kondo and Miura, 2010; Murray, 2003). An important property of the so-called Turing patterns is their dependency on domain size. If the size of the patterning domain is sufficiently increased or decreased further patterns appear or disappear respectively (Figure 3A). A biological example where such size dependency is observed is the marine angelfish Pomacanthus (Kondo and Asai, 1995). The coat of baby fish has few stripes, but more stripes form as the fish grows.

The Turing model was first applied to limb digit patterning in 1979 (Newman and Frisch, 1979), and many aspects of bone patterning in the limb have since been shown to be explicable with a Turing mechanism (Newman and Bhat, 2007). In particular, given their flexibility, Turing models could be shown to reproduce the wide range of different digit patterns on the various limb geometries of different species (Zhu et al., 2010). Miura and colleagues further showed that the supernumerary digits in the *Doublefoot* mutant mice (Crick et al., 2003) could be explained with the size dependency of Turing patterns (Miura et al., 2006). Newman and colleagues also showed that the different number of condensations in stylopod, zeugopod, and autopod in anterior-posterior direction could, in principle, result from the lengthening of the limb bud in proximal-distal direction; this effect requires the early fixation of proximal elements (Hentschel et al., 2004).

An increasing body of experimental results supports a Turing mechanism in the limb (Sheth et al., 2012). However, the molecular components of the Turing mechanism have remained elusive. Regulatory reactions, that would give rise to Turing patterns, were first defined about twenty years after the first publication of the Turing mechanism (Gierer and Meinhardt, 1972; Prigogine, 1967; Prigogine and Lefever, 1968). These included an activator-inhibitor mechanism and a so-called substrate-depletion mechanism. The activator-inhibitor mechanism has been frequently applied to reproduce patterns in biology because it provided an attractive regulatory framework that could be implemented by many negative feedback interactions in biological systems. One example is provided by the WNT-DKK negative feedback, which has been proposed to control hair follicle spacing (Sick et al., 2006). For the limb, a number of different negative feedbacks have been proposed, mainly based on TGF-β signalling and either fibronectin deposition in the extracellular matrix (ECM) or TGF-β antagonists (Zhu et al., 2010). However, to date there is no genetic evidence in support of any of these proposed Turing components. It has also been noted, that, the interaction with the ECM can result in self-emerging patterning. Thus by secreting enzymes, cells can digest the ECM and consecutively move closer. Such a traction-based mechanism then allows self-organized patterning (Oster, 1985). Similar to classical Turing pattern, haptotaxis can drive instabilities that can lead to the emergence of aggregation patterns, even in the absence of cell motility, i.e. without random movements of cells (Oster, 1983).

Part of the difficulty in proposing molecular components for Turing mechanisms is that one of the components (here the antagonist) has to diffuse much faster than the other (here the activator), which is difficult when both forms are diffusing. Transient differences in diffusion speeds as may result from differential interactions with the extracellular matrix have recently been suggested to result in Turing patterns (Müller et al., 2012). It will have to be seen whether such transient differences are indeed sufficient to give rise to robust symmetry breaks and patterning in biology. We have recently shown that different diffusion speeds as well as the other Turing conditions can easily be obtained with ligand-receptor pairs (Badugu et al., 2012; Celliere et al., 2012; Menshykau and Iber, 2013; Menshykau et al., 2012; Tanaka and Iber, 2013). Thus if ligands and receptors interact cooperatively and trigger the emergence of more receptor on the membrane (by enhancing expression or by enhancing receptor recycling to the membrane), these pairs give rise to Schnakenberg kinetics (Schnakenberg, 1979), which are well known to give rise to Turing patterns (Gierer and Meinhardt, 1972). This opens up the possibility that Turing patterns result from a single morphogen. We further showed that a model based on the BMP-receptor interaction would reproduce the digit patterning that is observed in wildtype and various mutants, both on static and growing domains, if we combined it with FGF signalling from the apical ectodermal ridge (Figure 3B,C) (Badugu et al., 2012). Interestingly, in *Smad4* mutants, *Sox9* is expressed in its characteristic horseshoe pattern in the autopod, but does not break up into digit condensations (Bénazet et al., 2012). SMAD4 is a CO-SMAD and thus a key transducer of canonical BMP signals.

In spite of the great similarity of simulated and real patterns, it remains to be established whether Turing-type mechanisms rather than alternative mechanisms underlie their establishment (Hofer and Maini, 1996). In fact, in several cases, Turing-type mechanisms have been wrongly assigned to patterning processes such as e.g. the mechanism by which the stripy expression pattern of pair-rule genes emerge during Drosophila development (Akam, 1989). These failures reveal the importance of a careful and comprehensive analysis of the underlying molecular interactions before proposing a Turing mechanism.

## Data-based mechanistic models of pattern formation in the limb

Experimental and genetic manipulation of vertebrate limb bud development over the last two decades has identified likely most key players controlling its growth and patterning, and has resulted in the definition of the core-regulatory network (Figure 4A). The regulatory network in the limb is complex and is further complicated by the spatially and temporally restricted expression of many of the components. To define the mechanism of digit patterning, it will be important to test and improve proposed mechanisms in a computational framework that is consistent with all solidly established experimental data.

Previous efforts to organize the vast knowledge about regulatory interactions in the limb typically resulted in a modular view, i.e. in the focus on small sub-networks. The sub-networks in the various publications of different mutant phenotypes are not always consistent with each other. An integrated analysis is possible only with the help of computational methods. Such more complex models are still best built by starting with a simple module that is then extended to include further factors that replace phenomenological descriptions in the model. In each step it is important to validate the model carefully with experimental data.

One way to organize the regulatory network into modules is to consider the control of the different axes separately and to only link these in a later step. Mouse forelimbs grow out from the flank and become visible around embryonic day (E) 9.25. As the limb bud grows and develops, asymmetries emerge along the proximal-distal (shoulder to fingertips), anterior-posterior (thumb to pinky), and dorso-ventral axes. The axes appear to be set up at different times and by different mechanisms. The proximal-distal (PD) axis develops as the limb bud expands distally as a result of the interaction between the apical ectodermal ridge (AER) and the underlying mesenchyme. The asymmetry in the anterior-posterior (AP) direction, on the other hand, is present already in the lateral plate mesenchyme as shown by graft experiments (Tabin, 1991). The dorso-ventral (DV) axis is established later and is defined by the ectoderm (MacCabe et al., 1974).

Limb bud outgrowth from the flank is initiated by a regulatory network of retinoic acid (RA), WNTs, and Fibroblastic Growth Factors (FGFs). Prior to limb initiation, *Fgf10* is expressed in a wide region in the trunk without any specific restriction to the presumptive limb areas. Around E8, RA induces the expression of *Wnt-2b* (Mercader et al., 2006), and WNT2B subsequently restricts the expression of *Fgf10* to the presumptive forelimb region; WNT8C plays a similar role for the hindlimb region (Kawakami et al., 2001). The inductive activities of both WNT2B and WNT8C are mediated by beta-catenin. Once limb initiation is underway, and after *Fgf10* expression has been restricted, FGF10 signals to the overlying ectoderm to induce expression of *Wnt-3a*, which eventually will become restricted to the AER. WNT3A then signals through beta-catenin to activate *Fgf8* expression. To complete the loop, FGF8 signals back to the mesenchyme of the nascent limb bud, where it contributes to maintain expression of *Fgf10*.

FGF8, together with the other FGFs that are expressed in the AER, control the distal part of the limb bud, while retinoic acid (RA) controls the proximal part of the limb bud (Figure 4B) (Cooper et al., 2011b; Mercader, 2000; Rosello-Diez et al., 2011). What leads to the separation of the two signalling centers? Genetic analysis in the mouse showed that FGF8 reduces the RA concentration by enhancing the expression of the RA metabolizing enzyme *Cyp26b1* in the distal mesenchyme (Probst et al., 2011). Mathematical simulations predicted that RA would in turn limit AER-FGF activity. Experimental analysis indeed confirmed that ectopic RA activity restricts *Fgf4* and to a lesser extent *Fgf8* expression in the AER (Figure 4C) (Probst et al., 2011). This revealed a mutually antagonistic interaction of RA with AER-FGFs. In line with this, earlier experiments had shown that low doses of an RA implant increase the length of the AER, while higher doses decrease the length of the ridge (Summerbell, 1983; Tickle et al., 1982). On the other hand, it has been noted that there is no proximal expansion in AER-*Fgf8* expression in conditional mutants of an RA producing enzyme (Cunningham et al., 2011). However, the same group also showed that RA is necessary for the initiation of forelimbs as otherwise ectopic *Fgf8* expression prevents forelimb initiation (Cunningham et al., 2013; Zhao et al., 2009). In the model, the antagonism between RA and FGF8 would be present already at the time of initiation, and the extension of the *Fgf8* domain could later well be maintained by RA-independent mechanisms. *Fgf8* expression is initiated upon limb but outgrowth in spite of the inhibitory role of RA, and the simulations predict that receptor binding limits diffusion of RA from the flank initially, once RA signalling enhances the expression of RA receptors, as indeed observed in experiments (Noji et al., 1991; Tabin, 1991). According to the model, receptor saturation eventually permits RA to diffuse further distally and to form a gradient that could regulate aspects of proximal-distal limb bud development (Figure 4C) and that could define the proximal part (stylopod -> humerus) of the proximal-distal axis as suggested by two recent studies (Cooper et al., 2011a; Roselló-Díez et al., 2011) and challenged by the Duester group (Cunningham et al., 2013; Zhao et al., 2009).

AER-FGFs initiate and maintain *Shh* expression, a key regulator of patterning along the anterior-posterior (AP) axis. SHH, in turn, enhances *AER-Fgf* expression via its impact on the expression of *Gremlin-1 (Grem1)*, an antagonist of BMPs (Bénazet and Zeller, 2009; Probst et al., 2011). The inclusion of such a simple positive feedback between AER-FGF and SHH in the mathematical model for the RA-AER-FGF interaction allowed us to reproduce the impact of a *Shh* knock-out on the patterning of the proximal-distal axis (Probst et al., 2011). The feedback between AER-FGFs and SHH thus integrates the development of the PD axis with that of the AP axis.

For a long time, it was unclear how *Shh* expression could be triggered by AER-FGFs in spite of a high BMP concentration, which has been shown to suppress *Shh* expression also in the presence of a high FGF concentration (Bastida et al., 2009). The conundrum was resolved by a combination of mathematical modelling and experimentation, which revealed that the BMP-dependent up-regulation of the BMP antagonist *Grem1* occurs fast (2h) (Bénazet et al., 2009). The resulting reduction in BMP activity enables up-regulation of *Shh* expression via AER-FGF signalling, which in turn enhances *Grem1* expression further. This robust and self-regulatory feedback signalling system propagates limb bud outgrowth distally and coordinates AP and PD limb bud axes development.

The initial simulation of the regulatory network of AER-FGF, SHH, BMP and GREM1 were only carried out over developmental time, but not on the spatial domain of the limb bud, and thus did not take the spatial differences along the different axes into account (Bénazet et al., 2009). The inclusion of the spatial domain is, however, important because it permits us to further test the consistency of proposed regulatory networks (Figure 4A,B) by comparing the spatio-temporal expression profiles of the network components to data from wild-type and mutant mice (Figure 4C). Such image-based comparisons can also be used to determine suitable parameter values for the model (Menshykau et al., 2013). Simulating such a rather complex network not only over time, but also over space on realistic 2D or 3D growing or static limb bud domains is numerically challenging, but feasible (Badugu et al., 2012; Germann et al. 2011; Iber et al., 2013; Menshykau and Iber, 2012). In developing the computational model, a number of inconsistencies in the verbal model are typically detected and resolved. Gaps in the understanding are highlighted that can be addressed in further experiments. A validated model finally allows the investigation of questions that are difficult to address by experiments alone. This concerns, in particular, functionalities that emerge indirectly from multiple regulatory interactions.

Going forward, the signalling models should be solved on realistic, growing domains to reveal how the self-organized regulation of the interaction network results in the emergence of digit condensations during limb bud outgrowth. Based on our previous model, the core network controlling the emergence of digit condensations comprises AER-FGFs and BMP (Figure 3) (Badugu et al., 2012), with SHH gradients from the ZPA likely determining digit identity. The BMP-receptor interaction can result in Turing pattern that mark the digit condensations, and AER-FGFs are necessary to move the Turing spots away from the boundary into the limb domain and to modulate the Turing patterns to obtain the correct wildtype and mutant digit patterns in the model (Badugu et al., 2012).

## Long Bone Development

To form digits, the digit condensations need to develop into the distinct long bones of the phalanges, separated by joints. The appearance of joints depends on the BMP antagonist NOGGIN, but long bones (without joints) still form in the absence of *Noggin* (Brunet et al., 1998). The first step in the formation of long bones is the formation of cartilage. The *Sox9*-expressing mesenchymal cells aggregate in mesenchymal condensations and subsequently develop into long bones by endochondral ossification (Figure 5A) (Kronenberg, 2003; Provot and Schipani, 2005; Wuelling and Vortkamp, 2011). As part of the process the digit condensations develop a highly organized spatial structure, the growth plate. In the developing growth plate, periarticular chondrocytes proliferate, differentiate into columnar chondrocytes, and then further differentiate into postmitotic hypertrophic chondrocytes (Figure 5A). Growth is a consequence of both proliferation and differentiation into larger hypertrophic cells. The volume increase upon differentiation into hypertrophic chondrocytes happens in three phases (Cooper et al., 2013). The first phase is characterized by true hypertrophy, i.e. a proportionate increase in dry mass production and fluid uptake. The enlargement in the second phase is the result of cell swelling and the dramatic dilution of cell dry mass. In the final third phase, cells increase in size by increasing both dry mass and fluid volume proportionally. Cell differentiation into hypertrophic chondrocytes, and at later stages apoptosis of hypertrophic chondrocytes and replacement by invading osteoblasts all start in the centre of the domain (Wuelling and Vortkamp, 2010; Wuelling and Vortkamp, 2011). Accordingly, hypertrophic chondrocytes (and later osteoblasts) accumulate at the centre of the domain, while proliferating (and resting) chondrocytes are found at the ends of the bone domain.

The study of mouse mutants has led to the identification of the core signalling proteins that control the growth and differentiation pattern during bone development by endochondral ossification (Figure 5B). A key signalling factor is Parathyroid hormone-related protein (PTHrP) (Lanske et al., 1996; Vortkamp et al., 1996). PTHrP increases the pool of mitotically active chondrocytes by preventing their differentiation into hypertrophic chondrocytes (Karaplis et al., 1994; Weir et al., 1996), but, unlike Hedgehog signaling, PTHrP does not enhance their proliferation rate (Karp et al., 2000). *Pthrp* is expressed and secreted by resting periarticular chondrocytes that reside at the ends of the domain (Karp et al., 2000), where differentiation is therefore blocked; hypertrophic chondrocytes thus emerge only at the centre of the domain. PTH/PTHrP-R signaling downregulates the expression of its own receptor, *Pth/Pthrp-r* (Kawane et al., 2003). *Pth/Pthrp* receptor (*Pth/Pthrp-r*) is therefore expressed in maturing chondrocytes and in the perichondrium/periosteum, i.e. in a zone adjacent to the post-mitotic prehypertrophic chondrocytes (Hilton et al., 2005; St-Jacques et al., 1999).

A second important regulator of endochondral ossification is Indian Hedgehog (IHH) (Kobayashi et al., 2005; Vortkamp et al., 1996). IHH signalling via its receptor PTCH1 induces the expression of *Pthrp* (Hilton et al., 2005; Karp et al., 2000; Kobayashi et al., 2005; St-Jacques et al., 1999; Vortkamp et al., 1996), as well as the expression of its own receptor, *Ptch1* (St-Jacques et al., 1999), and stimulates proliferation of chondrocytes (Figure 5B) (Karp et al., 2000). *Ptch1* is expressed most strongly in a zone adjacent to the post-mitotic pre-hypertrophic chondrocytes where also *Pth/Pthrp-r* is expressed (Hilton et al., 2005). In the *Ihh* null mouse no expression of *Ptch1* and *Pthrp* is observed in the developing bone, *Pth/Pthrp-r* is misexpressed, the chondrocyte proliferation rate is lower, and only very few hypertrophic chondrocytes emerge in the center of the domain (St-Jacques et al., 1999). IHH signals also independently of PTHrP. Thus IHH acts on periarticular chondrocytes to stimulate their differentiation, thereby regulating the columnar cell mass (Kobayashi et al., 2005), and further promotes chondrocyte hypertrophy (Mak et al., 2008). PTH/PTHrP-R signalling downregulates the action of IHH (Kobayashi et al., 2005). High levels of *Ihh* mRNA are detected in cartilage from as early as E11.5 (Bitgood and McMahon, 1995). Expression is highest in chondrocytes in the growth regions of developing bones, but a lower level of expression persists into the hypertrophic zone (Bitgood and McMahon, 1995). On maturation, expression becomes progressively restricted to post-mitotic pre-hypertrophic chondrocytes adjacent to the *Pth/Pthrp-r* -expressing proliferative zones (Bitgood and McMahon, 1995; Vortkamp et al., 1996).

While the core regulatory network, comprising PTHrP, IHH, and its receptor PTCH1, has been defined, it has remained unclear how the patterns and the spatio-temporal control of the process emerge from these interactions. A number of mathematical models have been developed to explain the distribution of the signalling proteins IHH and PTHrP and their impact on bone growth and development (Bougherara et al., 2010; Brouwers et al., 2006; Garzon-Alvarado et al., 2009; Garzon-Alvarado et al., 2010; Isaksson et al., 2008; van Donkelaar and Huiskes, 2007). Garzon-Alvarado and colleagues suggest that regulatory interactions between IHH and PTHrP result in Schnakenberg-like reaction kinetics (Garzon-Alvarado et al., 2009), which can give rise to Turing pattern (Gierer and Meinhardt, 1972). In particular, they postulate that the rate of PTHrP production and IHH removal are both proportional to the concentration of PTHrP squared times the IHH concentration ($[PTHrP]^2 [IHH]$). While IHH indeed enhances *Pthrp* expression (Karsenty et al., 2009), PTHrP signaling negatively impacts on its own expression (Kobayashi et al., 2005), which contradicts a key assumption of the model. Moreover, there is no experimental evidence that PTHrP would enhance IHH turn-over; PTHrP rather blocks *Ihh* production by preventing hypertrophic differentiation (Vortkamp et al., 1996) and downregulates the action of IHH (Kobayashi et al., 2005). The reaction kinetics in the model are thus unlikely to reflect the physiological situation.

We had previously shown that the interaction of the Hedgehog protein with its receptor PTCH1 can result in a Schnakenberg-type Turing mechanism and that this mechanism can explain the observed patterning dynamics during lung branching morphogenesis (Menshykau et al., 2012). Unlike in the lung where an increasing number of branches appear as the structure is growing out, the number of patterns during endochondral ossification, however, does not increase (except for the late emergence of the secondary ossification center in each end (epiphysis) of the long bones). Moreover, as a result of cell differentiation during bone development, the production rate of proteins changes continuously. This is a challenge in any kind of Turing mechanism as Turing patterns typically arise only within a very small parameter range, the Turing space. We nonetheless found that a model that couples the IHH-

PTCH1-based Schnakenberg-type Turing mechanism with the underlying tissue dynamics could still generate the observed patterns on a growing and differentiating tissue domain, i.e. the emergence of hypertrophic chondrocytes and *Ihh* expression in the centre of the domain, the predominance of proliferating chondrocytes towards the sides of the domain, and the emergence of a differentiation zone towards the centre of the domain) (Tanaka and Iber, 2013). The inclusion of PTHrP was important to achieve robust patterning when coupling patterning and growth.

In spite of the good match of simulations and embryonic patterning dynamics, the model has two important limitations. For one, patterning only works for growth speeds that are similar to those observed in the mouse, while higher growth speeds as they may be present in jerboa (Cooper et al., 2013) would be difficult to accommodate. Moreover, while the model reproduced most mutant phenotypes it failed to explain the normal early patterning that is observed in the *Ihh*$^{-/-}$; *Gli3*$^{-/-}$ double knock-out (Hilton et al., 2005). Both in the wildtype and in the double knock-out *Pthrp* expression is restricted to the ends of the domain (Hilton et al., 2005). It thus seems that the main role of IHH signalling is to supress the repressor action of the transcription factor GLI3 (Hilton et al., 2005), and that in the absence of IHH/GLI3 signalling there is an alternative patterning process that restricts *Pthrp* expression to the sides and thereby restricts the emergence of hypertrophic chondrocytes to the centre.

One possible mechanism, by which *Pthrp* expression may be restricted to the ends of the domain, are factors that are secreted by the joints. Various ligands from the TGF-β family are all present in the joints and SMAD3-dependent signalling has previously been shown to stimulate *Pthrp* expression (Pateder et al., 2001; Pateder et al., 2000). In line with this, BMP receptor BMPR-IA (ALK3) signalling induces *Ihh* and *Pthrp* expression, and expression of a constitutive active form results in pattern reversal, similar to that observed when *Pthrp* is overexpressed (Zou et al., 1997). Similarly, constitutive active ALK2 in the developing chick limb bud induces *Ihh* and *Pthrp* expression and delays the formation of hypertrophic chondrocytes (Zhang et al., 2003). BMP-4 and GDF-5 bind to activin receptor-like kinase 3 (ALK-3) and/or ALK-6 (also termed BMP type IA and type IB receptors, respectively), whereas BMP-6 and BMP-7 preferentially bind to ALK-2 (Aoki et al., 2001). In the E14.5 limbs only *Alk-3/BmprIa* is found to be expressed in columnar proliferating and early hypertrophic chondrocytes, while the expression of all other BMP receptors is restricted to the perichondrium and bone (Minina et al., 2005). BMPs/GDF-5 in the joints may thus induce *Pthrp* expression in the perichondrium and *Ihh* expression in the pre-hypertrophic chondrocytes. BMP ligands signal through various receptors, but all canonical BMP signalling requires the CO-SMAD SMAD4. Early *Prx1-Cre*-mediated conditional inactivation of *Smad4* in the limb bud mesenchyme results in defects upstream of endochondral ossification and lack of collagen type II synthesis, a marker of proliferating maturing chondrocytes (Bénazet et al., 2012); in the *Prx1-Cre*-mediated mutant *Smad4* transcripts are absent already at E9.5 (Bénazet et al., 2012). Later conditional removal of *Smad4* with *Hoxa13-Cre* still permits *Sox9* expression in the digit ray primordia, but, interestingly, no cartilage or ossification is observed in the autopod (Bénazet et al., 2012). TGF-β ligands, on the other hand, are no good candidates to stimulate *Pthrp* expression as conditional removal of the TGFβ-type-II receptor (*TβrII*) in the limb enhances (rather than reduces) *Pthrp* expression (Longobardi et al., 2012; Spagnoli et al., 2007).

In summary, expression of *Pthrp* during endochondral ossification may be regulated by both BMP and IHH signalling. Both signalling pathways can be described by Schnakenberg-like reaction kinetics (Badugu et al., 2012; Menshykau et al., 2012), which can give rise to Turing pattern (Gierer and Meinhardt, 1972). Once *Pthrp* expression has been restricted to the outer parts of the condensations, the emergence of hypertrophic chondrocytes naturally becomes restricted to the center of the condensations, resulting in the characteristic pattern that is observed during endochondral ossification.

# Conclusion

Biological functionality is largely controlled by sophisticated networks of interacting proteins. Much is known about the interactions that control organ growth and patterning, but until very recently, this information was largely presented in verbal models and cartoons. Due to the inherent complexity, experiments, in general, only validate small modules but not larger, integrated models. With the help of computational models it has recently become possible to integrate the available data in larger frameworks that begin to provide insights into apparently counterintuitive experimental data sets, detect inconsistencies in models and datasets, and have predictive power for new informative experiments. Such an integrative approach relies on careful experimental validation of all key elements of these models and simulations. Following their experimental validation, these models can be used for *in silico* genetics, i.e. their predictive power allows simulations of mutant states and the resulting phenotypes in situations when experimental generation and analysis of e.g. compound mutant embryos would be difficult or very time consuming. This will allow better focusing of research on the relevant genetic experiments and avoid the generation of uninformative mutant embryos. On the other hand, match and/or discrepancies between *in silico* and real genetics will reveal and/or improve the validity of the current mechanistic model. We illustrated this approach by the on-going development of *in silico* simulations of limb bud and digit development. Numerical solutions of these patterning processes are challenging, but will become progressively more feasible with advances in algorithms and computing power.

# Acknowledgments

The authors thank Rolf Zeller and Erkan Ünal for critical comments on the manuscript and acknowledge funding from the SNF Sinergia grant "Developmental engineering of endochondral ossification from mesenchymal stem cells".

**REFERENCES**

**Akam, M.** (1989). Drosophila development: making stripes inelegantly. In *Nature*, pp. 282-283.

**Amano, T., Sagai, T., Tanabe, H., Mizushina, Y., Nakazawa, H. and Shiroishi, T.** (2009). Chromosomal dynamics at the Shh locus: limb bud-specific differential regulation of competence and active transcription. *Dev Cell* **16**, 47-57.

**Aoki, H., Fujii, M., Imamura, T., Yagi, K., Takehara, K., Kato, M. and Miyazono, K.** (2001). Synergistic effects of different bone morphogenetic protein type I receptors on alkaline phosphatase induction. *Journal of cell science* **114**, 1483-1489.

**Badugu, A., Kraemer, C., Germann, P., Menshykau, D. and Iber, D.** (2012). Digit patterning during limb development as a result of the BMP-receptor interaction. *Scientific reports* **2**, 991.

**Bastida, M. F., Sheth, R. and Ros, M. A.** (2009). A BMP-Shh negative-feedback loop restricts Shh expression during limb development. *Development (Cambridge, England)* **136**, 3779-3789.
**Ben-Zvi, D., Pyrowolakis, G., Barkai, N. and Shilo, B.-Z.** (2011). Expansion-repression mechanism for scaling the Dpp activation gradient in Drosophila wing imaginal discs. *Current biology : CB* **21**, 1391-1396.
**Ben-Zvi, D., Shilo, B.-Z., Fainsod, A. and Barkai, N.** (2008). Scaling of the BMP activation gradient in Xenopus embryos. *Nature* **453**, 1205-1211.
**Bénazet, J.-D., Bischofberger, M., Tiecke, E., Gonçalves, A., Martin, J. F., Zuniga, A., Naef, F. and Zeller, R.** (2009). A self-regulatory system of interlinked signaling feedback loops controls mouse limb patterning. *Science* **323**, 1050-1053.
**Bénazet, J.-D., Pignatti, E., Nugent, A., Unal, E., Laurent, F. and Zeller, R.** (2012). Smad4 is required to induce digit ray primordia and to initiate the aggregation and differentiation of chondrogenic progenitors in mouse limb buds. *Development (Cambridge, England)* **139**, 4250-4260.
**Bénazet, J.-D. and Zeller, R.** (2009). Vertebrate limb development: moving from classical morphogen gradients to an integrated 4-dimensional patterning system. *Cold Spring Harbor perspectives in biology* **1**, a001339.
**Bitgood, M. J. and McMahon, A. P.** (1995). Hedgehog and Bmp genes are coexpressed at many diverse sites of cell-cell interaction in the mouse embryo. *Developmental Biology* **172**, 126-138.
**Boehm, B., Westerberg, H., Lesnicar-Pucko, G., Raja, S., Rautschka, M., Cotterell, J., Swoger, J. and Sharpe, J.** (2010). The role of spatially controlled cell proliferation in limb bud morphogenesis. *PLoS Biol* **8**, e1000420.
**Bollenbach, T., Pantazis, P., Kicheva, A., Bökel, C., Gonzalez-Gaitan, M. and Julicher, F.** (2008). Precision of the Dpp gradient. *Development (Cambridge, England)* **135**, 1137-1146.
**Bougherara, H., Klika, V., Marsik, F., Marik, I. A. and Yahia, L. a. H.** (2010). New predictive model for monitoring bone remodeling. *Journal Of Biomedical Materials Research Part A* **95**, 9-24.
**Brouwers, J. E. M., van Donkelaar, C. C., Sengers, B. G. and Huiskes, R.** (2006). Can the growth factors PTHrP, Ihh and VEGF, together regulate the development of a long bone? *Journal of biomechanics* **39**, 2774-2782.
**Brunet, L. J., McMahon, J. A., McMahon, A. P. and Harland, R. M.** (1998). Noggin, cartilage morphogenesis, and joint formation in the mammalian skeleton. *Science* **280**, 1455-1457.
**Celliere, G., Menshykau, D. and Iber, D.** (2012). Simulations demonstrate a simple network to be sufficient to control branch point selection, smooth muscle and vasculature formation during lung branching morphogenesis. *Biology Open*, 1-14.
**Chiang, C., Litingtung, Y., Lee, E., Young, K. E., Corden, J. L., Westphal, H. and Beachy, P. A.** (1996). Cyclopia and defective axial patterning in mice lacking Sonic hedgehog gene function. *Nature* **383**, 407-413.
**Cooper, K. L., Hu, J. K.-H., ten Berge, D., Fernandez-Teran, M., Ros, M. A. and Tabin, C. J.** (2011a). Initiation of proximal-distal patterning in the vertebrate limb by signals and growth. *Science* **332**, 1083-1086.

**Cooper, K. L., Hu, J. K. H., ten Berge, D., Fernandez-Teran, M., Ros, M. A. and Tabin, C. J.** (2011b). Initiation of Proximal-Distal Patterning in the Vertebrate Limb by Signals and Growth. *Science* **332**, 1083-1086.

**Cooper, K. L., Oh, S., Sung, Y., Dasari, R. R., Kirschner, M. W. and Tabin, C. J.** (2013). Multiple phases of chondrocyte enlargement underlie differences in skeletal proportions. *Nature* **495**, 375-378.

**Crick, A. P., Babbs, C., Brown, J. M. and Morriss-Kay, G. M.** (2003). Developmental mechanisms underlying polydactyly in the mouse mutant Doublefoot. *Journal of anatomy* **202**, 21-26.

**Cunningham, T. J., Chatzi, C., Sandell, L. L., Trainor, P. A. and Duester, G.** (2011). Rdh10 mutants deficient in limb field retinoic acid signaling exhibit normal limb patterning but display interdigital webbing. *Developmental dynamics : an official publication of the American Association of Anatomists* **240**, 1142-1150.

**Cunningham, T. J., Zhao, X., Sandell, L. L., Evans, S. M., Trainor, P. A. and Duester, G.** (2013). Antagonism between Retinoic Acid and Fibroblast Growth Factor Signaling during Limb Development. *Cell reports*.

**Dillon, R. and Othmer, H. G.** (1999). A mathematical model for outgrowth and spatial patterning of the vertebrate limb bud. *Journal of theoretical biology* **197**, 295-330.

**Ede, D. A. and Law, J. T.** (1969). Computer simulation of vertebrate limb morphogenesis. *Nature* **221**, 244-248.

**Garzon-Alvarado, D. A., Garcia-Aznar, J. M. and Doblare, M.** (2009). A reaction-diffusion model for long bones growth. *Biomechanics And Modeling In Mechanobiology* **8**, 381-395.

**Garzon-Alvarado, D. A., Peinado Cortes, L. M. and Cardenas Sandoval, R. P.** (2010). A mathematical model of epiphyseal development: hypothesis of growth pattern of the secondary ossification centre. *Computer methods in biomechanics and biomedical engineering*, 1.

**Germann, P., Menshykau, D., Tanaka, S. and Iber, D.** (2011) Simulating Organogenesis with Comsol *Proceedings of COMSOL Conference 2011*.

**Gierer, A. and Meinhardt, H.** (1972). A theory of biological pattern formation. *Kybernetik* **12**, 30-39.

**Gregor, T., Tank, D. W., Wieschaus, E. F. and Bialek, W.** (2007). Probing the limits to positional information. *Cell* **130**, 153-164.

**Harfe, B. D., Scherz, P. J., Nissim, S., Tian, H., McMahon, A. P. and Tabin, C. J.** (2004). Evidence for an expansion-based temporal Shh gradient in specifying vertebrate digit identities. *Cell* **118**, 517-528.

**Hentschel, H. G. E., Glimm, T., Glazier, J. A. and Newman, S. A.** (2004). Dynamical mechanisms for skeletal pattern formation in the vertebrate limb. *Proc Biol Sci* **271**, 1713-1722.

**Hilton, M. J., Tu, X., Cook, J., Hu, H. and Long, F.** (2005). Ihh controls cartilage development by antagonizing Gli3, but requires additional effectors to regulate osteoblast and vascular development. *Development (Cambridge, England)* **132**, 4339-4351.

**Hofer, T. and Maini, P.** (1996). Turing patterns in fish skin? *Nature* **380**, 678.

**Horton, W. A. and Degnin, C. R.** (2009). FGFs in endochondral skeletal development. *Trends in endocrinology and metabolism: TEM* **20**, 341-348.

**Iber, D., Tanaka, S., Fried, P., Germann, P. and Menshykau, D.** (2013). Simulating Tissue Morphogenesis and Signaling In *Tissue Morphogenesis: Methods and Protocols* (ed. C. M. Nelson): Methods in Molecular Biology (Springer).

**Iber, D. and Zeller, R.** (2012). Making sense-data-based simulations of vertebrate limb development. *Curr Opin Genet Dev* **22**, 570-577.

**Isaksson, H., van Donkelaar, C. C., Huiskes, R. and Ito, K.** (2008). A mechano-regulatory bone-healing model incorporating cell-phenotype specific activity. *Journal of theoretical biology* **252**, 230-246.

**Karaplis, A. C., Luz, A., Glowacki, J., Bronson, R. T., Tybulewicz, V. L., Kronenberg, H. M. and Mulligan, R. C.** (1994). Lethal skeletal dysplasia from targeted disruption of the parathyroid hormone-related peptide gene. *Genes Dev* **8**, 277-289.

**Karp, S. J., Schipani, E., St-Jacques, B., Hunzelman, J., Kronenberg, H. and McMahon, A. P.** (2000). Indian hedgehog coordinates endochondral bone growth and morphogenesis via parathyroid hormone related-protein-dependent and -independent pathways. *Development (Cambridge, England)* **127**, 543-548.

**Karsenty, G., Kronenberg, H. M. and Settembre, C.** (2009). Genetic control of bone formation. *Annu Rev Cell Dev Biol* **25**, 629-648.

**Kawakami, Y., Capdevila, J., Büscher, D., Itoh, T., Rodriguez-Esteban, C. and Izpisúa Belmonte, J. C.** (2001). WNT signals control FGF-dependent limb initiation and AER induction in the chick embryo. *Cell* **104**, 891-900.

**Kawakami, Y., Tsuda, M., Takahashi, S., Taniguchi, N., Esteban, C. R., Zemmyo, M., Furumatsu, T., Lotz, M., Belmonte, J. C. I. and Asahara, H.** (2005). Transcriptional coactivator PGC-1alpha regulates chondrogenesis via association with Sox9. *Proc. Natl. Acad. Sci USA* **102**, 2414-2419.

**Kawane, T., Mimura, J., Yanagawa, T., Fujii-Kuriyama, Y. and Horiuchi, N.** (2003). Parathyroid hormone (PTH) down-regulates PTH/PTH-related protein receptor gene expression in UMR-106 osteoblast-like cells via a 3',5'-cyclic adenosine monophosphate-dependent, protein kinase A-independent pathway. *The Journal of endocrinology* **178**, 247-256.

**Kobayashi, T., Soegiarto, D. W., Yang, Y., Lanske, B., Schipani, E., McMahon, A. P. and Kronenberg, H. M.** (2005). Indian hedgehog stimulates periarticular chondrocyte differentiation to regulate growth plate length independently of PTHrP. *The Journal of clinical investigation* **115**, 1734-1742.

**Kondo and Asai** (1995). A reaction–diffusion wave on the skin of the marine angelfish Pomacanthus. *Nature* **376**, 765-768.

**Kondo, S. and Miura, T.** (2010). Reaction-diffusion model as a framework for understanding biological pattern formation. *Science* **329**, 1616-1620.

**Kraus, P., Fraidenraich, D. and Loomis, C. A.** (2001). Some distal limb structures develop in mice lacking Sonic hedgehog signaling. *Mechanisms of development* **100**, 45-58.

**Kronenberg, H. M.** (2003). Developmental regulation of the growth plate. *Nature* **423**, 332-336.

**Lander, A. D., Lo, W.-C., Nie, Q. and Wan, F. Y. M.** (2009). The Measure of Success: Constraints, Objectives, and Tradeoffs in Morphogen-mediated Patterning. *Cold Spring Harbor perspectives in biology* **1**, a002022.
**Lanske, B., Karaplis, A. C., Lee, K., Luz, A., Vortkamp, A., Pirro, A., Karperien, M., Defize, L. H., Ho, C., Mulligan, R. C., et al.** (1996). PTH/PTHrP receptor in early development and Indian hedgehog-regulated bone growth. *Science* **273**, 663-666.
**Lauschke, V. M., Tsiairis, C. D., Francois, P. and Aulehla, A.** (2013). Scaling of embryonic patterning based on phase-gradient encoding. *Nature* **493**, 101-105.
**Litingtung, Y., Dahn, R. D., Li, Y., Fallon, J. F. and Chiang, C.** (2002). Shh and Gli3 are dispensable for limb skeleton formation but regulate digit number and identity. *Nature* **418**, 979-983.
**Longobardi, L., Li, T., Myers, T. J., O'Rear, L., Ozkan, H., Li, Y., Contaldo, C. and Spagnoli, A.** (2012). TGF-β type II receptor/MCP-5 axis: at the crossroad between joint and growth plate development. *Dev Cell* **23**, 71-81.
**MacCabe, J. A., Errick, J. and Saunders, J. W.** (1974). Ectodermal control of the dorsoventral axis in the leg bud of the chick embryo. *Developmental Biology* **39**, 69-82.
**Maini, P. K. and Solursh, M.** (1991). Cellular mechanisms of pattern formation in the developing limb. *International review of cytology* **129**, 91-133.
**Mak, K. K., Kronenberg, H. M., Chuang, P.-T., Mackem, S. and Yang, Y.** (2008). Indian hedgehog signals independently of PTHrP to promote chondrocyte hypertrophy. *Development (Cambridge, England)* **135**, 1947-1956.
**Menshykau, D. and Iber, D.** (2012). Simulating Organogenesis with Comsol: Interacting and Deforming Domains *Proceedings of COMSOL Conference 2012*.
---- (2013). Kidney branching morphogenesis under the control of a ligand-receptor-based Turing mechanism. *Physical biology* **10**, 046003.
**Menshykau, D., Kraemer, C. and Iber, D.** (2012). Branch Mode Selection during Early Lung Development. *Plos Computational Biology* **8**, e1002377.
**Menshykau, D., Shrivastsan, A., Germann, P., Lemereux, L. and Iber, D.** (2013). Simulating Organogenesis in COMSOL: Parameter Optimization for PDE-based models. In *Proceedings of COMSOL Conference 2013, Rotterdam*.
**Mercader, N., Fischer, S. and Neumann, C. J.** (2006). Prdm1 acts downstream of a sequential RA, Wnt and Fgf signaling cascade during zebrafish forelimb induction. *Development (Cambridge, England)* **133**, 2805-2815.
**Mercader, N., Leonardo, E., Piedra, M., Martinez-A, C., Ros, M. A., Torres, M. S.** (2000). Opposing RA and FGF signals control proximodistal vertebrate limb development through regulation of Meis genes. *Development* **127**, 3961–3970.
**Minina, E., Schneider, S., Rosowski, M., Lauster, R. and Vortkamp, A.** (2005). Expression of Fgf and Tgfbeta signaling related genes during embryonic endochondral ossification. *Gene expression patterns : GEP* **6**, 102-109.
**Miura, T., Shiota, K., Morriss-Kay, G. and Maini, P. K.** (2006). Mixed-mode pattern in Doublefoot mutant mouse limb--Turing reaction-diffusion model on a growing domain during limb development. *Journal of theoretical biology* **240**, 562-573.

**Müller, P., Rogers, K. W., Jordan, B. M., Lee, J. S., Robson, D., Ramanathan, S. and Schier, A. F.** (2012). Differential diffusivity of Nodal and Lefty underlies a reaction-diffusion patterning system. *Science* **336**, 721-724.

**Murray, J. D.** (2003). *Mathematical Biology. 3rd edition in 2 volumes: Mathematical Biology: II. Spatial Models and Biomedical Applications.*: Springer.

**Newman, S. A. and Bhat, R.** (2007). Activator-inhibitor dynamics of vertebrate limb pattern formation. *Birth defects research Part C, Embryo today : reviews* **81**, 305-319.

**Newman, S. A. and Frisch, H. L.** (1979). Dynamics of skeletal pattern formation in developing chick limb. *Science* **205**, 662-668.

**Noji, S., Nohno, T., Koyama, E., Muto, K., Ohyama, K., Aoki, Y., Tamura, K., Ohsugi, K., Ide, H. and Taniguchi, S.** (1991). Retinoic acid induces polarizing activity but is unlikely to be a morphogen in the chick limb bud. *Nature* **350**, 83-86.

**Oster, G., Murray, J, Maini, P** (1985). A model for chondrogenic condensations in the developing limb the role of extracellular matrix and cell tractions. *Journal of embryology and experimental morphology* **112**, 93-112.

**Oster, G., Murray, JD, Harris, AK** (1983). Mechanical aspects of mesenchymal morphogenesis. *Journal of embryology and experimental morphology* **78**, 83-125.

**Pateder, D. B., Ferguson, C. M. and Ionescu, A. M.** (2001). PTHrP expression in chick sternal chondrocytes is regulated by TGF‐$\beta$ through Smad‐mediated signaling. *Journal of cellular ⋯*.

**Pateder, D. B., Rosier, R. N., Schwarz, E. M., Reynolds, P. R., Puzas, J. E., D'Souza, M. and O'Keefe, R. J.** (2000). PTHrP expression in chondrocytes, regulation by TGF-beta, and interactions between epiphyseal and growth plate chondrocytes. *Experimental cell research* **256**, 555-562.

**Prigogine, I.** (1967). On Symmetry-Breaking Instabilities in Dissipative Systems. *J Chem Phys* **46**, 3542-3550.

**Prigogine, I. and Lefever, R.** (1968). Symmetry Breaking Instabilities in Dissipative Systems. II. *The Journal of Chemical Physics* **48**, 1695.

**Probst, S., Kraemer, C., Demougin, P., Sheth, R., Martin, G. R., Shiratori, H., Hamada, H., Iber, D., Zeller, R. and Zuniga, A.** (2011). SHH propagates distal limb bud development by enhancing CYP26B1-mediated retinoic acid clearance via AER-FGF signalling. *Development (Cambridge, England)* **138**, 1913-1923.

**Provot, S. and Schipani, E.** (2005). Molecular mechanisms of endochondral bone development. *Biochemical and biophysical research communications* **328**, 658-665.

**Riddle, R. D., Johnson, R. L., Laufer, E. and Tabin, C.** (1993). Sonic hedgehog mediates the polarizing activity of the ZPA. *Cell* **75**, 1401-1416.

**Rosello-Diez, A., Ros, M. A. and Torres, M.** (2011). Diffusible Signals, Not Autonomous Mechanisms, Determine the Main Proximodistal Limb Subdivision. *Science* **332**, 1086-1088.

**Roselló-Díez, A., Ros, M. A. and Torres, M.** (2011). Diffusible signals, not autonomous mechanisms, determine the main proximodistal limb subdivision. *Science* **332**, 1086-1088.

**Saunders, J. W. and Gasseling, M. T.** (1968). Ectoderm-mesenchymal interactions in the origin of wing symmetry. . In *Epithelial-Mesenchymal Interactions.* (ed. F. R & B. RE), pp. 78-97: Baltimore: Williams and Wilkins.

**Schnakenberg, J.** (1979). Simple chemical reaction systems with limit cycle behaviour. *Journal of theoretical biology* **81**, 389-400.

**Sheth, R., Marcon, L., Bastida, M. F., Junco, M., Quintana, L., Dahn, R., Kmita, M., Sharpe, J. and Ros, M. A.** (2012). Hox genes regulate digit patterning by controlling the wavelength of a Turing-type mechanism. *Science* **338**, 1476-1480.

**Sick, S., Reinker, S., Timmer, J. and Schlake, T.** (2006). WNT and DKK determine hair follicle spacing through a reaction-diffusion mechanism. *Science* **314**, 1447-1450.

**Spagnoli, A., O'Rear, L., Chandler, R. L., Granero-Molto, F., Mortlock, D. P., Gorska, A. E., Weis, J. A., Longobardi, L., Chytil, A., Shimer, K., et al.** (2007). TGF-beta signaling is essential for joint morphogenesis. *J Cell Biol* **177**, 1105-1117.

**St-Jacques, B., Hammerschmidt, M. and McMahon, A. P.** (1999). Indian hedgehog signaling regulates proliferation and differentiation of chondrocytes and is essential for bone formation. *Genes Dev* **13**, 2072-2086.

**Summerbell, D.** (1983). The effect of local application of retinoic acid to the anterior margin of the developing chick limb. *Journal of embryology and experimental morphology* **78**, 269-289.

**Tabin, C. J.** (1991). Retinoids, homeoboxes, and growth factors: toward molecular models for limb development. *Cell* **66**, 199-217.

**Tanaka, S. and Iber, D.** (2013). Inter-dependent tissue growth and Turing patterning in a model for long bone development. *Physical biology* **10**.

**te Welscher, P., Zuniga, A., Kuijper, S., Drenth, T., Goedemans, H. J., Meijlink, F. and Zeller, R.** (2002). Progression of vertebrate limb development through SHH-mediated counteraction of GLI3. *Science* **298**, 827-830.

**Tickle, C.** (1981). The number of polarizing region cells required to specify additional digits in the developing chick wing. *Nature* **289**, 295-298.

**Tickle, C., Alberts, B., Wolpert, L. and Lee, J.** (1982). Local application of retinoic acid to the limb bond mimics the action of the polarizing region. *Nature* **296**, 564-566.

**Tickle, C., Shellswell, G., Crawley, A. and Wolpert, L.** (1976). Positional signalling by mouse limb polarising region in the chick wing bud. *Nature* **259**, 396-397.

**Turing, A.** (1952). The chemical basis of morphogenesis. *Phil. Trans. Roy. Soc. Lond* **B237**, 37-72.

**Umulis, D. M.** (2009). Analysis of dynamic morphogen scale invariance. *J R Soc Interface* **6**, 1179-1191.

**Umulis, D. M., Shimmi, O., O'Connor, M. B. and Othmer, H. G.** (2010). Organism-Scale Modeling of Early Drosophila Patterning via Bone Morphogenetic Proteins. *Developmental Cell* **18**, 260-274.

**van Donkelaar, C. C. and Huiskes, R.** (2007). The PTHrP-Ihh feedback loop in the embryonic growth plate allows PTHrP to control hypertrophy and Ihh to regulate proliferation. *Biomechanics And Modeling In Mechanobiology* **6**, 55-62.

**Vortkamp, A., Lee, K., Lanske, B., Segre, G. V., Kronenberg, H. M. and Tabin, C. J.** (1996). Regulation of rate of cartilage differentiation by Indian hedgehog and PTH-related protein. *Science* **273**, 613-622.

**Weir, E. C., Philbrick, W. M., Amling, M., Neff, L. A., Baron, R. and Broadus, A. E.** (1996). Targeted overexpression of parathyroid hormone-related peptide in chondrocytes causes chondrodysplasia and delayed endochondral bone formation. *Proceedings of the National Academy of Sciences of the United States of America* **93**, 10240-10245.

**Wolpert, L.** (1969). Positional information and the spatial pattern of cellular differentiation. *Journal of theoretical biology* **25**, 1-47.

**Wuelling, M. and Vortkamp, A.** (2010). Transcriptional networks controlling chondrocyte proliferation and differentiation during endochondral ossification. *Pediatric nephrology (Berlin, Germany)* **25**, 625-631.

---- (2011). Chondrocyte proliferation and differentiation. *Endocrine development* **21**, 1-11.

**Yang, Y., Drossopoulou, G., Chuang, P. T., Duprez, D., Martí, E., Bumcrot, D., Vargesson, N., Clarke, J., Niswander, L., McMahon, A., et al.** (1997). Relationship between dose, distance and time in Sonic Hedgehog-mediated regulation of anteroposterior polarity in the chick limb. *Development (Cambridge, England)* **124**, 4393-4404.

**Zeller, R., López-Ríos, J. and Zuniga, A.** (2009). Vertebrate limb bud development: moving towards integrative analysis of organogenesis. *Nat Rev Genet* **10**, 845-858.

**Zhang, D., Schwarz, E. M., Rosier, R. N., Zuscik, M. J., Puzas, J. E. and O'Keefe, R. J.** (2003). ALK2 functions as a BMP type I receptor and induces Indian hedgehog in chondrocytes during skeletal development. *Journal of bone and mineral research : the official journal of the American Society for Bone and Mineral Research* **18**, 1593-1604.

**Zhao, X., Sirbu, I. O., Mic, F. A., Molotkova, N., Molotkov, A., Kumar, S. and Duester, G.** (2009). Retinoic acid promotes limb induction through effects on body axis extension but is unnecessary for limb patterning. *Current biology : CB* **19**, 1050-1057.

**Zhu, J., Nakamura, E., Nguyen, M.-T., Bao, X., Akiyama, H. and Mackem, S.** (2008). Uncoupling Sonic hedgehog control of pattern and expansion of the developing limb bud. *Dev Cell* **14**, 624-632.

**Zhu, J., Zhang, Y.-T., Alber, M. S. and Newman, S. A.** (2010). Bare bones pattern formation: a core regulatory network in varying geometries reproduces major features of vertebrate limb development and evolution. *PLoS ONE* **5**, e10892.

**Zou, H., Wieser, R., Massagué, J. and Niswander, L.** (1997). Distinct roles of type I bone morphogenetic protein receptors in the formation and differentiation of cartilage. *Genes Dev* **11**, 2191-2203.

**FIGURES**

**Figure 1: Digit Emergence during Limb Development**

(A) Nascent digit condensations become first visible around embryonic day (E) 11.5 and express *Sox9* (Kawakami et al., 2005). (B) The digits appear neither simultaneously nor in strict posterior to anterior sequence (or vice-versa) (Zhu et al., 2008).

**Figure 2: The French Flag Model**

(A) A threshold-based differentiation mechanism. A morphogen diffuses from the left into the domain, and the cells differentiate according to two fixed thresholds into three cell types, illustrated by the colours of the French flag. (B) A lower morphogen concentration would result in less digits (or digit types) in the French Flag model as the high threshold concentration would not be breached. (C) The impact of changes in the morphogen source. Already very small changes in the morphogen source would shift the differentiation fronts. (D) The impact of domain size. The limb buds of different species differ in size, and the pattern would not scale relative to the size of the limb domain if based on a simple French Flag mechanism, i.e. blue and white regions are of normal size, while the red region is extended.

**Figure 3: Turing Pattern as a Mechanism for Digit Patterning**

(A) Interactions between receptors (R) and ligands (L) can give rise to Schnakenberg kinetics that result in Turing patterns. To result in Turing patterns, ligands need to diffuse faster than receptors, receptors and ligands need to bind cooperatively, and receptor-ligand binding must upregulate the receptor density on the membrane. An important property of Turing patterns is their dependency on domain size. If the size of the patterning domain is sufficiently increased or decreased further patterns appear or disappear respectively. (B, C) Digit patterning in a ligand-receptor based Turing model based on BMP, BMP receptor, that is modulated by AER-FGFs. (B) The regulatory network proposed in (Badugu et al., 2012). AER-FGFs are produced in the AER, marked in red, and regulate BMP expression in the AER and mesenchyme. The dimer BMP, B, and its receptor, R, form a complex composed of one BMP dimer and two receptors, $BR^2$. (C) The BMP-receptor, $BR^2$, first forms the characteristic donut shape that then breaks into digit condensations as the limb bud is growing out; simulations were carried out on a domain that grows at the measured speed.

**Figure 4: Data-based Models of Pattern Formation in the Limb**

(A) The regulatory network including the most important molecular players, i.e. RA, BMP (expressed in the green domain), FGF secreted from the AER (red line), SHH secreted from the ZPA (yellow) and GREM1 (expression domain marked in purple). (B) The regulatory sub-network that controls the PD axis. AER-FGFs induce expression of the enzyme *Cyp26b1*, which promotes the turn-over of RA. RA, in turns, represses expression of AER-*Fgf*s. (C) Patterning the PD axis. Simulation of the RA-AER-FGF network with (1st column) and without (2nd column) the negative regulation of AER-*Fgf* expression by RA (red arrow in panel B). Both simulations reproduce the observed restriction of *Rarb* expression (1st row) to the proximal side of the limb bud (3rd column). However, only in case of the negative impact of RA on AER-FGF (2nd column) can the distal restriction AER-*Fgf* expression be reproduced (2nd row). As predicted by the model, RA-loaded beads indeed downregulate AER-*Fgf* expression (3rd column) (Probst et al., 2011).

**Figure 5: The control of Long Bone Formation**

(A) Endochondral bone development from mesenchymal condensation stage to formation of mature growth plate. (a, b) Chondrocytes differentiate within mesenchymal condensations to form cartilage anlagen of future bones. (c, d) Coincident with the appearance of the perichondrial bone collar, chondrocytes in the central anlage hypertrophy followed by invasion of vascular and osteoblastic cells from the collar (e) and formation of the primary ossification center (f). This process expands toward the ends of the bone, eventually forming mature growth plates (h). Secondary ossification centers later form in the epiphyseal cartilage (i). With permission, from (Horton and Degnin, 2009). (B) A cartoon of the cell types, the differentiation paths, and the regulatory network controlling long bone formation by endochondral ossification. For details see text.

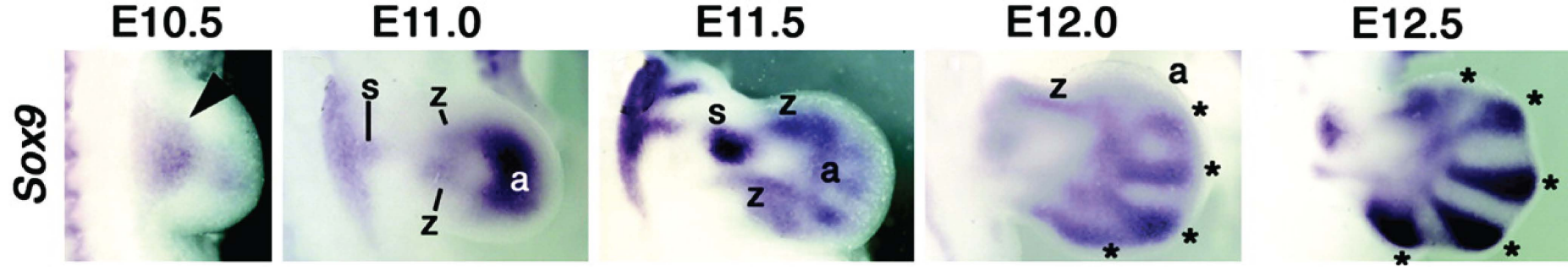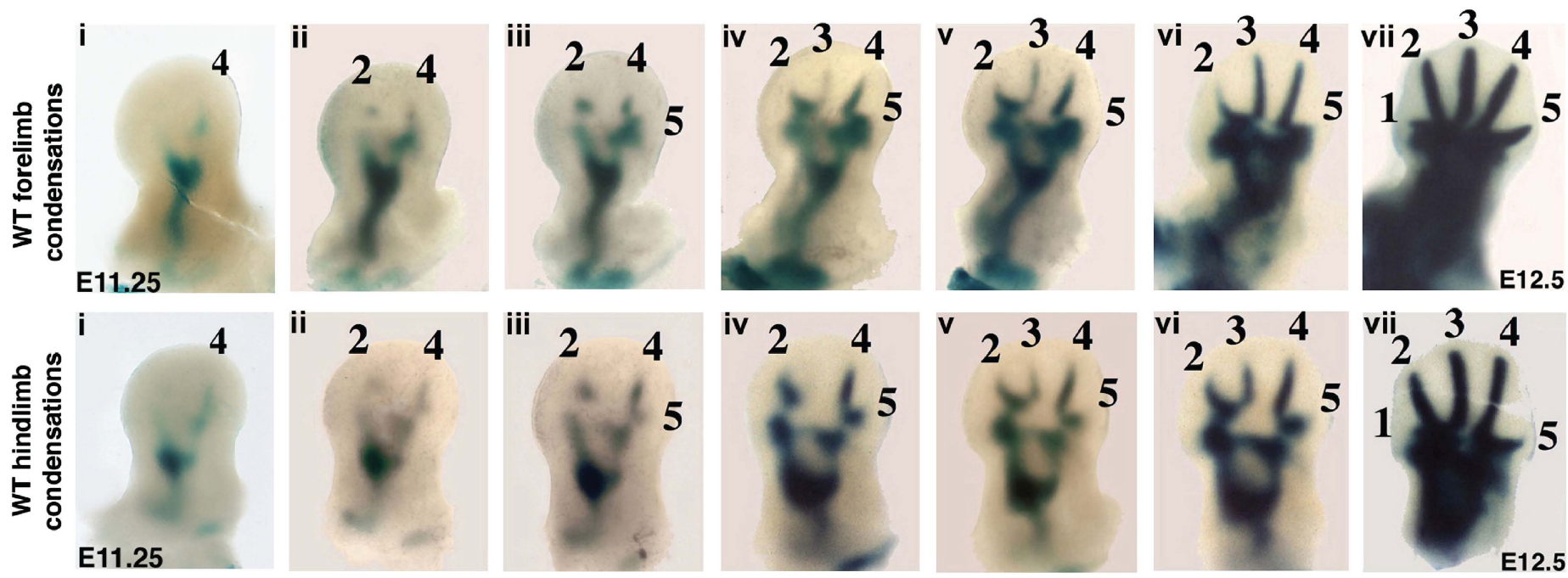

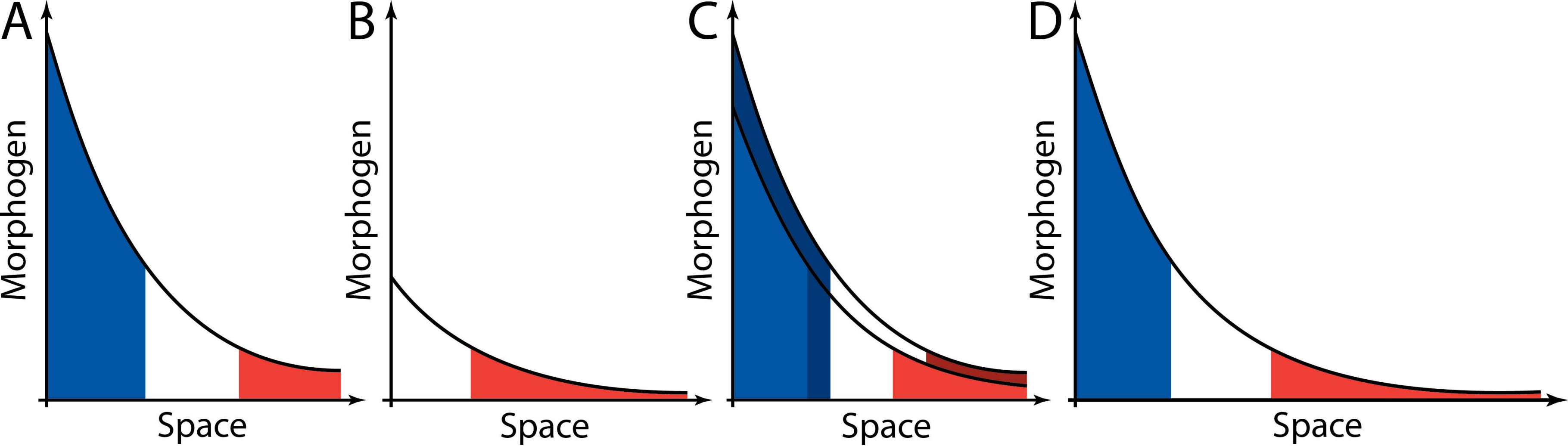

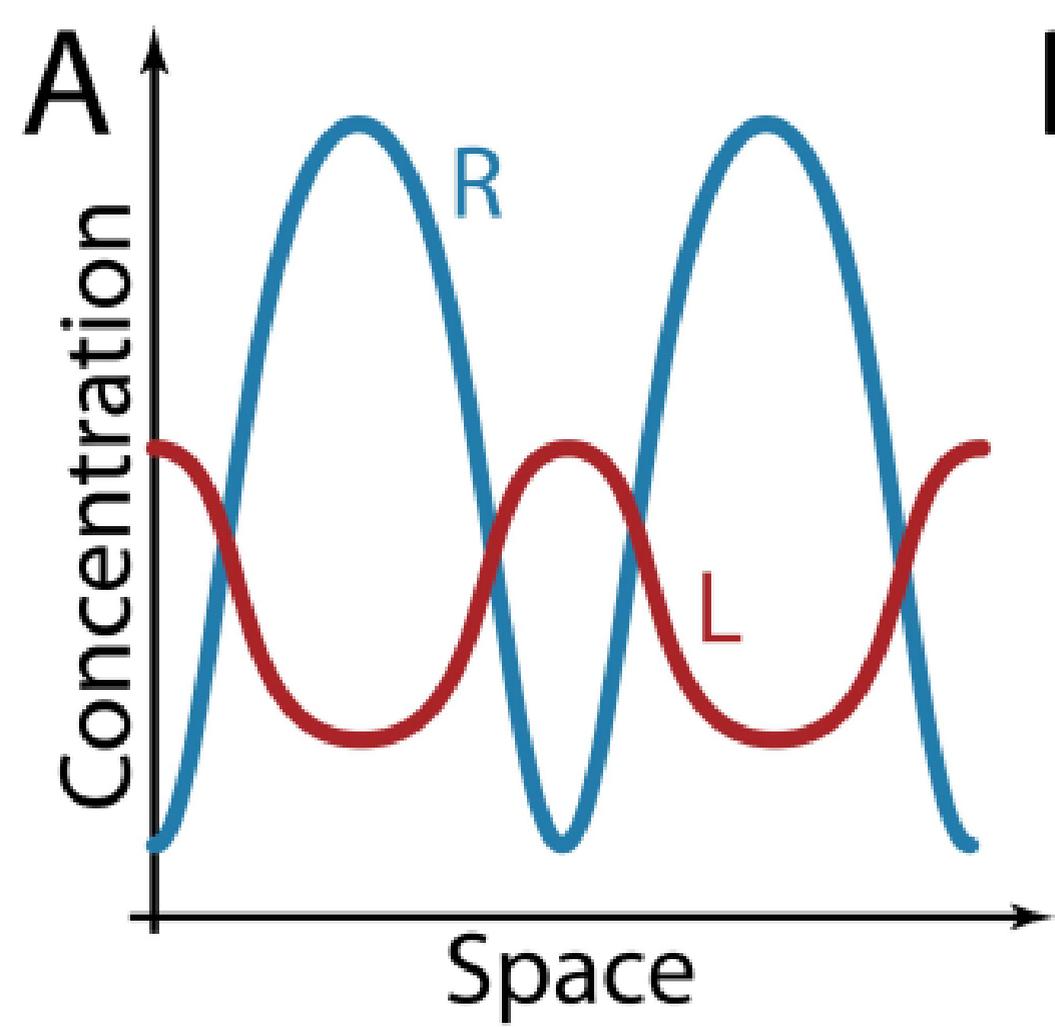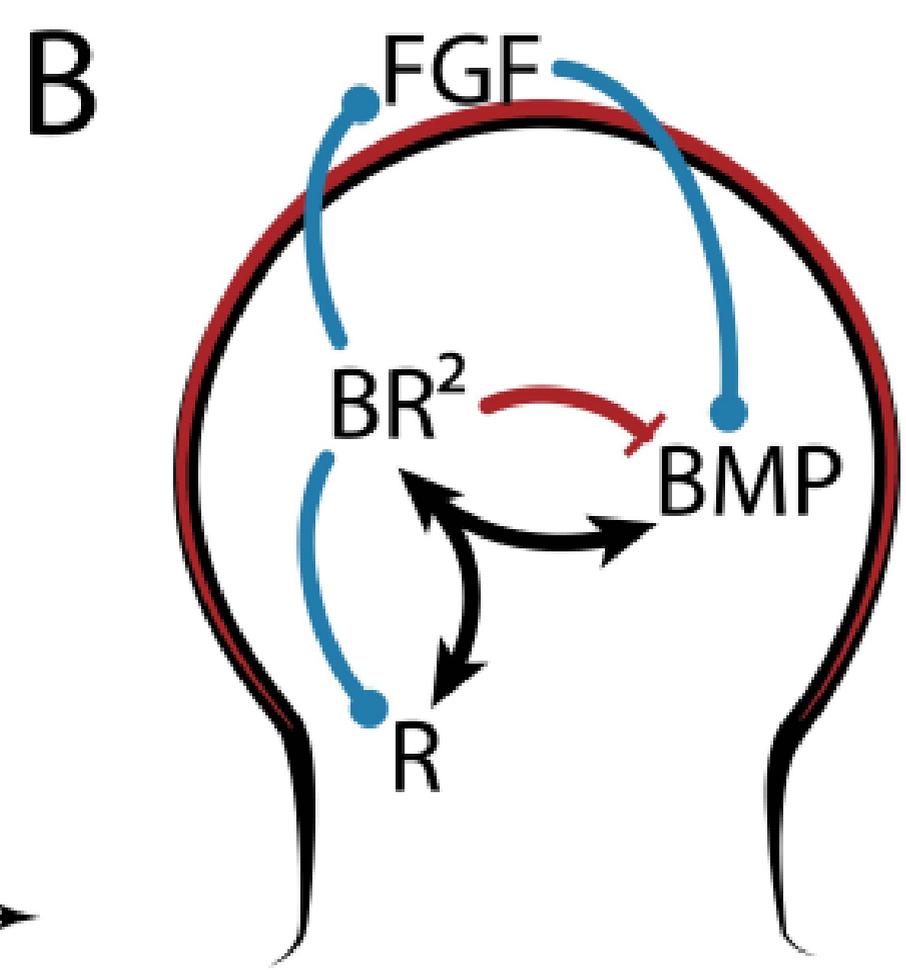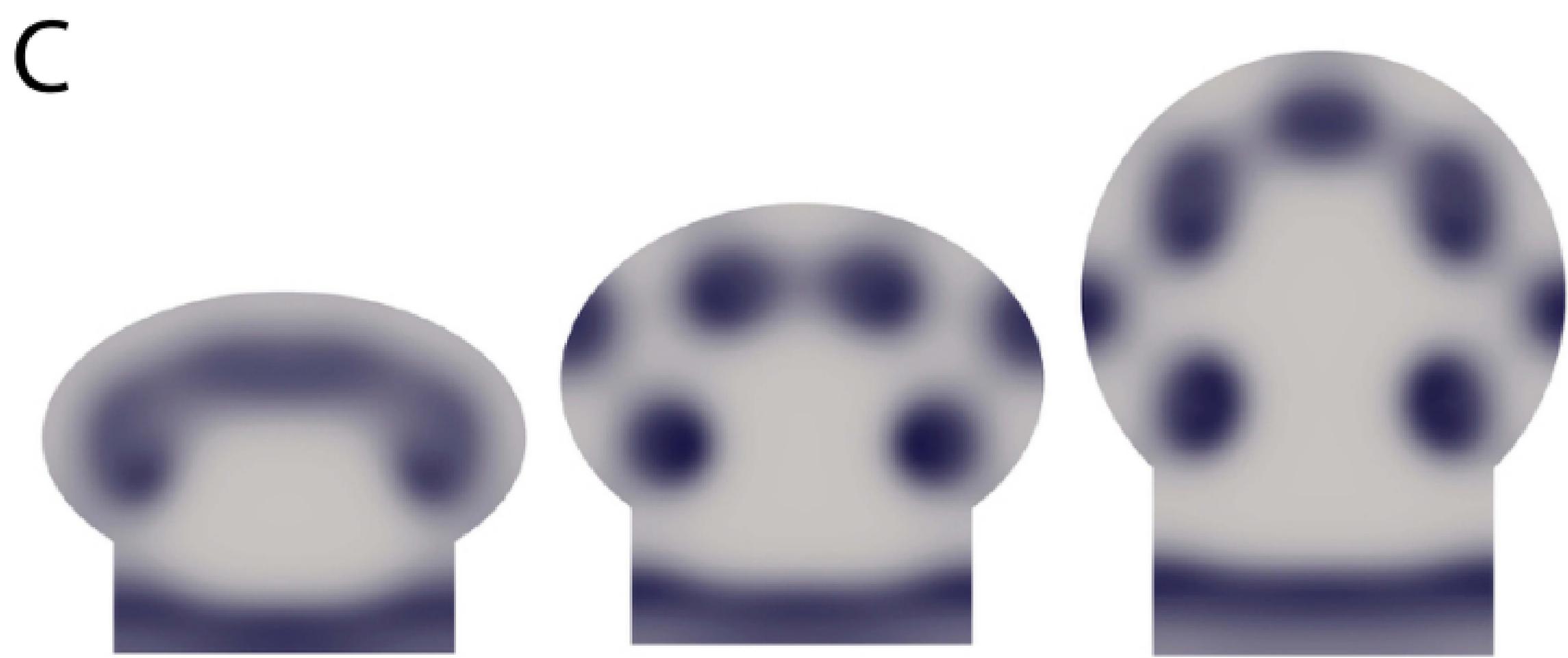

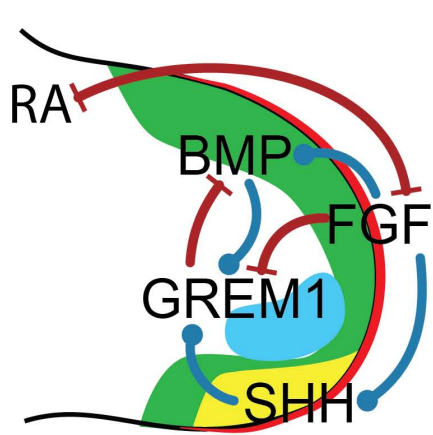
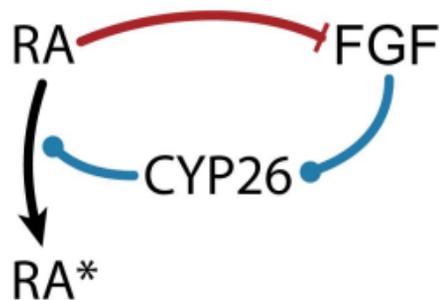
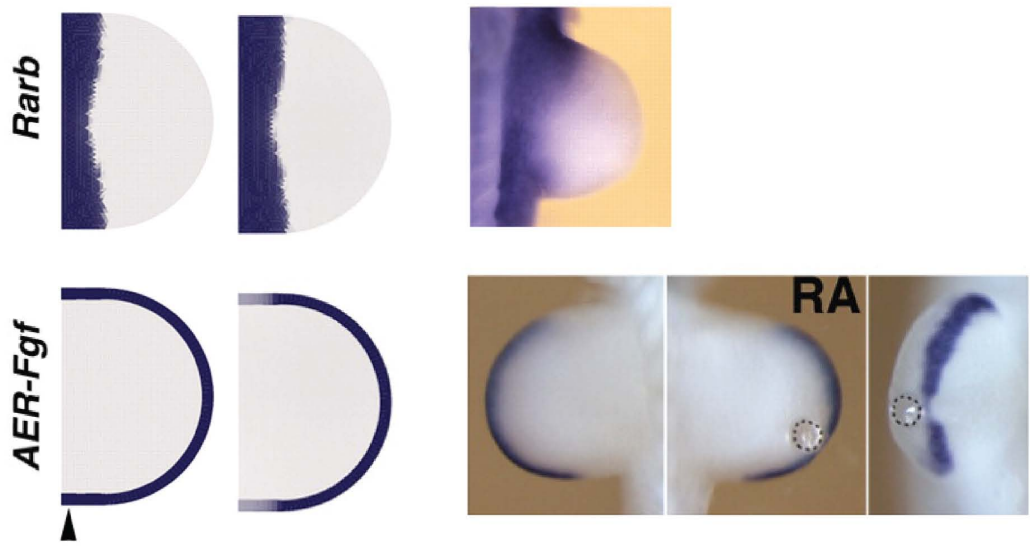

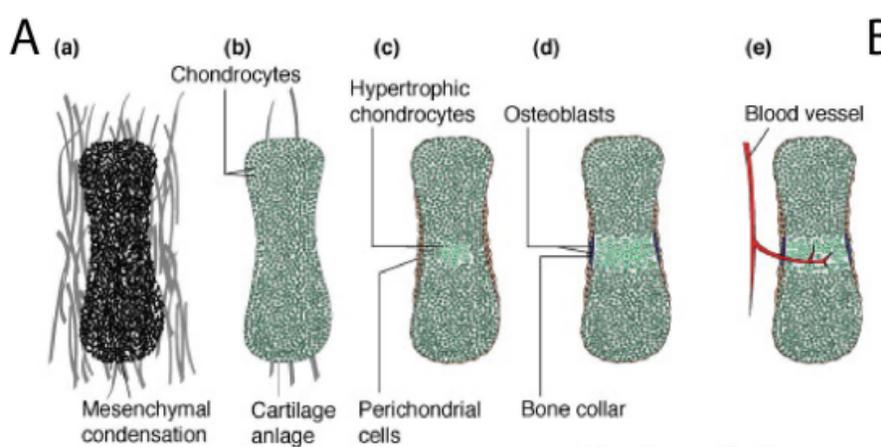
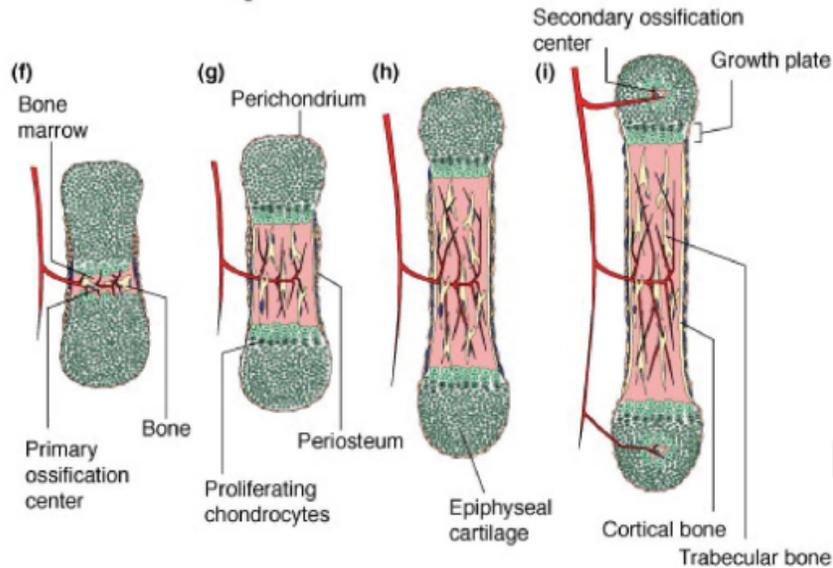
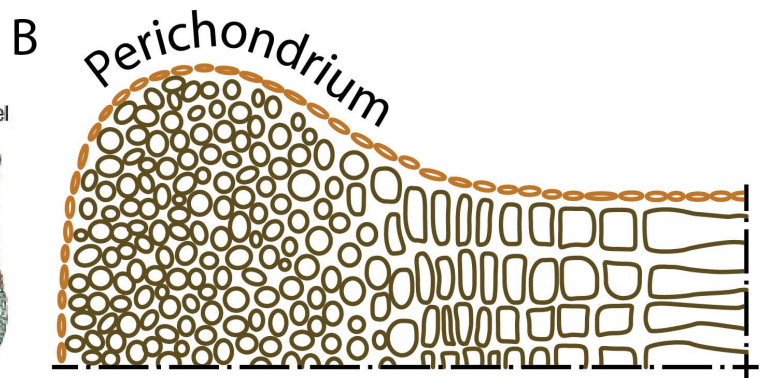
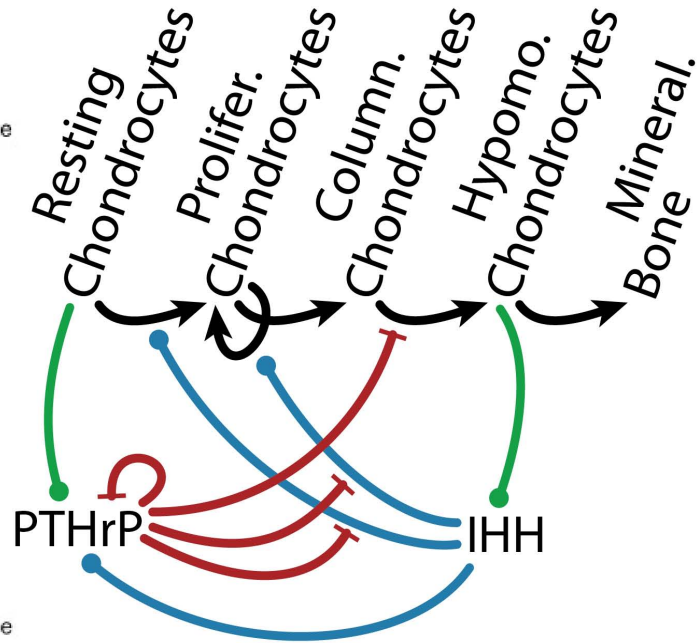